\documentclass[letterpaper, 10 pt, conference]{ieeeconf} 
\IEEEoverridecommandlockouts                              
\usepackage{ntheorem}
\usepackage{amsmath}
\usepackage{amsfonts}
\usepackage{authblk}
\newtheorem{theorem}{Theorem}

\newtheorem{lemma}{Lemma}

\newtheorem{assumption}{Assumption}

\usepackage{listings}
\usepackage{color}
\usepackage{array}
\usepackage{tabu}
\usepackage{cite}
\usepackage{mathrsfs,amssymb}
\usepackage{graphicx}
\graphicspath{ {image22/} }
\usepackage{float}
\usepackage{algorithm}
\usepackage{algpseudocode}
\DeclareMathOperator*{\argmin}{arg\,min}

\DeclareMathOperator*{\card}{card}
\DeclareMathOperator*{\supp}{supp}

\usepackage{algorithm}
\usepackage{algpseudocode}
\usepackage{mathrsfs,amssymb}
\usepackage{graphicx}
\usepackage{float}
\graphicspath{ {nuio/} }
\usepackage{listings}
\usepackage{color}
\usepackage{array}
\usepackage{tabu}
\DeclareMathOperator*{\rank}{rank}
\begin{document}
	%
	\title{On Joint Reconstruction of State and Input-Output Injection Attacks for Nonlinear Systems}
	%
	%
	
	\author{Tianci Yang \textsuperscript{1}, Carlos Murguia \textsuperscript{2}, Chen Lv\textsuperscript{1}, Dragan Ne\v{s}i\'{c}\textsuperscript{3}, and Chao Huang\textsuperscript{1}
		\thanks{This work was supported by the SUG-NAP Grant (No. M4082268.050) of Nanyang Technological University, Singapore}
		\thanks{\textsuperscript{1} The authors are with the School of Mechanical and Aerospace Engineering, Nanyang Technological University, Singapore. Emails:
			\{tianci.yang, lyuchen, chao.huang\}@ntu.edu.sg}%
		\thanks{\textsuperscript{2} The author is with the Department of Mechanical Engineering, Eindhoven University of Technology, The Netherlands. Email:
			c.g.murguia@tue.nl}
		\thanks{\textsuperscript{3} The author is with the Department of Electrical and Electronic Engineering, University of Melbourne, Australia. Email:
			dnesic@unimelb.edu.au}%
	}
	\maketitle
	
	\begin{abstract}
		We address the problem of robust state reconstruction for discrete-time nonlinear systems when the actuators and sensors are injected with (potentially unbounded) attack signals. Exploiting redundancy in sensors and actuators and using a bank of unknown input observers (UIOs), we propose an observer-based estimator capable of providing asymptotic estimates of the system state and attack signals under the condition that the numbers of sensors and actuators under attack are sufficiently small. Using the proposed estimator, we provide methods for isolating the compromised actuators and sensors. Numerical examples are provided to demonstrate the effectiveness of our methods.
	\end{abstract}
	

	%
	\IEEEpeerreviewmaketitle

	\section{Introduction}
	%
	%
	%
	%
	
	Networked Control Systems (NCSs) are firmly embedded in many aspects of our daily lives. Compared with traditional control systems, NCSs bring a number of advanges such as low installation/maintenance cost, reduced weight/volume, remote diagnosis/control. Recently, security of NCSs has become a main concern as wireless communication networks might serve as new access points for malicious agents trying to deteriorate the functionality of systems. According to the 2015 US security report, the number of of cyberattacks on critical infrastructure has increased 3,000\% from 2009 to 2015. In particular, transportation, energy and water has been their main target, which might lead to catastrophic fatalities, financial loss, and threaten national security. It follows that we need strategic mechanisms for attack identification and mitigation on NCSs.
	
	In \cite{Fawzi2012}-\nocite{Massoumnia1986}\nocite{Pajic2014}\nocite{Mo2014}\nocite{Vamvoudakis2014}\nocite{Chong2016b}\nocite{Vamvoudakis2012}\nocite{Shoukry2014}\nocite{Yong}\nocite{Park2015}\nocite{Liu2009}\nocite{Teixeira2012b}\nocite{Murguia2016}\nocite{Dolk1}\nocite{Hashemil2017}\nocite{Pasqualetti123}\nocite{Jairo}\nocite{murguia2020security}\nocite{Sahand2017}\cite{ahmed2017model}, several security-related research problems for linear control systems have been investigated. In general, analysis tools are provided for quantifying how different classes of cyberattacks degrade system performance; reaction strategies are given to identify and mitigate their effect on the system dynamics. There are quite few results addressing the nonlinear case, although many engineering systems are nonlinear in nature. In \cite{Kim2016}, the authors design algorithms for sensor attack detection and state estimation for uniformly observable continuous-time nonlinear systems. In \cite{hu2017secure}, the authors provide a compressed sensing-based estimation algorithm for nonlinear power systems under sensor attacks. In \cite{yang2020multi}, we provide an estimation framework for general nonlinear systems under sensor attacks.
	
	In this manuscript, we extend our idea in \cite{yang2020unknown}, where an unknown input multi-observer estimator is designed for linear time-invariant systems under sensor and actuator attacks, and consider using UIOs as a tool to solve the problem of secure estimation, attack reconstruction and isolation for discrete-time nonlinear systems in the presence of sensor and actuator attacks.
	We first consider the
	case when the system has $n_{u}$ actuators and all of which are potentially 
	attacked by an adversary and only a subset of its $n_{y}$ sensors are under attack. Using a bank of complete UIOs as the main ingredient, we construct an estimator capable of providing robust state estimates independent of the actuator and sensor attack signals. The main idea of designing the estimator is the following. Each complete UIO in the bank assumes all inputs are unavailable and is driven by different subsets of sensors. Thus, if the sensors are attack-free, these complete UIOs produce stable estimation errors. For every pairs of complete UIOs, we compute the largest difference between their estimates. If a pair of complete UIOs are driven by healthy sensors, then these complete UIOs produce the smallest difference between their estimates and provide good estimates of the system states. Then, we assume complete UIOs are not available, however, partial UIOs which can estimate the system states when some inputs are unknown exist and only a subset of actuators and sensors are under attack. We use a bank of partial UIOs as the main ingredient to construct an estimator capable of providing robust estimates of the system state despite the occurrence of actuator and sensor attacks. The main idea of our approach is as follows. Each partial UIO in the bank assumes a different set of inputs are unavailable and is driven by different subsets of sensors. Thus, if the inputs assumed to be unknown by the UIOs include all the attacked ones and the sensors are attack-free, these UIOs produce attack-free estimates. We compute the largest difference between the estimates given by every pair of partial UIOs in the bank. If the inputs assumed to be unknown by a pair of UIOs include all the attacked ones and the sensor measurements they use for estimation are attack-free, then these UIOs produce the smallest difference between their estimates and provide good estimates of the system states. Next, we propose a method for isolating false data injection actuator and sensor attacks once an estimate of the system state is obtained. 
	
	The paper is organized as follows. In Section \ref{estimation}, two types of UIO-based estimators are given. In Section \ref{isolation}, a method for isolating actuator and sensor attacks are proposed. Illustrative examples are presented to illustrate the performance of the estimators, the method of isolating attacks. Finally, we give concluding remarks in Section \ref{conclusion}.
	
	\textit{Notations}:
	We denote the set of real numbers by $\mathbb{R}$, the set of natural numbers by $\mathbb{N}$, and $\mathbb{R}^{n\times m}$ the set of $n\times m$ matrices for any $m,n \in \mathbb{N}$. For any vector $v\in\mathbb{R}^{n}$,  we denote {$v_{J}$} the stacking of all $v_{i}$, $i\in J$, $J\subset \left\lbrace 1,\hdots,n\right\rbrace$, $|v|=\sqrt{v^{\top} v}$, and $\supp(v)=\left\lbrace i\in\left\lbrace 1,\hdots,n\right\rbrace |v_{i}\neq0\right\rbrace $. We denote the cardinality of a set $S$ as $\card(S)$. The binomial coefficient is denoted as $\binom{a}{b}$, where $a,b$ are nonnegative integers. We denote a variable $m$ uniformly distributed in the interval $(a,b)$ as $m\sim\mathcal{U}(a,b)$. A continuous function $\alpha:[0,a)]\to[0,\infty)]$ is said to belong to class $\mathcal{K}$ if it is strictly increasing and $\alpha(0)=0$ \cite{khalil2002nonlinear}. Similarly, a continuous function $\beta(r,s)$ belongs to class $\mathcal{KL}$ if, for fixed $s$, the mapping $\beta(r,s)$ belongs to class $\mathcal{K}$ with respect to $r$ and, for fixed $r$, the mapping $\beta(r,s)$ is decreasing with respect to $s$ and $\beta(r,s)\to 0$ as $s\to\infty$ \cite{khalil2002nonlinear}.
	\section{Estimation}\label{estimation}
	Consider a discrete-time nonlinear system under sensor and actuator attacks:
	\begin{eqnarray}\label{c3s1}
		\begin{split}
			x^{+}=&f(x)+B(u+a_{u}),\\
			y=&h(x)+a_{y},
		\end{split}
	\end{eqnarray}
	where state $x\in\mathbb{R}^{n}$, output $y\in\mathbb{R}^{n_{y}}$, known input $u\in\mathbb{R}^{n_{u}}$, vector of actuator attacks $a_{u}\in\mathbb{R}^{n_{u}}, a_{u}=(a_{u1},\ldots,a_{un_{u}})^{\top}$, i.e., $a_{ui}(k)=0$ for all $k\geq 0$ if the $i$-th actuator is attack-free; otherwise, $a_{ui}(k_{i})\neq 0$ for some $k_{i}\geq 0$ and can be arbitrarily large, and vector of sensor attacks $a_{y}\in\mathbb{R}^{n_{y}}, a_{y}=(a_{y1},\ldots,a_{yn_{y}})^{\top}$, i.e., $a_{yi}(k)=0$ for all $k\geq 0$ if the $i$-th sensor is attack-free; otherwise, $a_{yi}(k_{i})\neq 0$ for some $k_{i}\geq 0$ and can be arbitrarily large. Let $B$ have full column rank, $W_{u}\subseteq\left\lbrace 1,\hdots,n_{u}\right\rbrace $ denote the \emph{unknown} set of actuators under attack, and $W_{y}\subset\left\lbrace 1,\ldots,n_{y}\right\rbrace $ be the \emph{unknown} set of sensors under attack. We assume the following.
	\begin{assumption}\label{c3asump2}
		The sets of attacked actuators and sensors remain constant over time, i.e., $W_{u}\subset\left\lbrace 1,\ldots,n_{u}\right\rbrace ,W_{y}\subset\left\lbrace 1,\ldots,n_{y}\right\rbrace $ are constant (time-invariant) and $\supp(a_{u}(k))\subseteq W_{u}$,\linebreak $\supp(a_{y}(k))\subseteq W_{y}$, for all $k\geq 0$.
	\end{assumption}
	
	\subsection{Complete Unknown Input Observers}\label{c3complete}
	We first treat $a_{u}$ in \eqref{c3s1} as an unknown input to system (\ref{c3s1}) and consider an observer of the form:
	\begin{equation}\label{c3o}
		\hat{x}_{J_{s}}^{+}=f_{J_{s}}(\hat{x}_{J_{s}},u,y^{J_{s}},(y^{J_{s}})^{+}),
	\end{equation}
	where $\hat{x}_{J_{s}}$ is the observer state and $f_{J_{s}}:\mathbb
	{R}^{n}\times\mathbb{R}^{n_{u}}\times\mathbb{R}^{\card(J_{s})}\times\mathbb{R}^{\card(J_{s})}\to\mathbb{R}^{n}$ denotes some nonlinear function. Define $e_{J_s}=\hat{x}_{J_s}-x$. System \eqref{c3o} is a complete unknown input observer for system \eqref{c3s1} if, for all $a_{u}\in\mathbb{R}^{n_{u}}$, and $a_{y}^{J_{s}}(k)=0$, $\forall k\geq 0$, there exist a $\mathcal{KL}$-function $\beta_{J_s}(\cdot,\cdot)$, such that:
	\begin{equation}\label{c3error}
		|e_{J_s}(k)|\leq \beta_{J_s}(|e_{J_s}(0)|,k),
	\end{equation}
	for all $e_{J_{s}}(0)\in\mathbb{R}^{n}$ and $k\geq 0$.
	
	Let $q$ be the largest integer such that for each $y^{J_s}\in\mathbb{R}^{\card(J_{s})}$ with $J_{s}\subset\left\lbrace 1,\ldots,n_{y}\right\rbrace $ and $\card(J_{s})\geq n_{y}-2q>0$, a complete UIO of the form \eqref{c3o} satisfying \eqref{c3error} exists.
	
	\begin{assumption}\label{c3a1}
		There are at most $q$ sensors attacked by an adversary, i.e.,
		\begin{equation}
			\card(W_{y})\leq q<\frac{n_{y}}{2},
		\end{equation}
		where $q$ denotes the largest integer such that for all $J_{s}\subset\left\lbrace 1,\ldots,n_{y}\right\rbrace$ with $\card(J_{s})\geq n_{y}-2q$, a complete UIO \eqref{c3o} exists for any $y^{J_{s}}\in\mathbb{R}^{\card(J_{s})}$.
	\end{assumption}
	
	
	\begin{lemma}\label{c3l4}
		If Assumption \ref{c3a1} is satisfied, among each set of $n_{y}-q$ sensors, at least $n_{y}-2q>0$ of them are attack-free.
	\end{lemma}
	\emph{\textbf{Proof:}} Lemma \ref{c3l4} follows trivially from Assumption \ref{c3a1}.\hfill$\blacksquare$\\[2mm]
	Suppose a complete UIO is designed for each set $J_{s}\subset\left\lbrace 1,\ldots,n_{y}\right\rbrace $ with $\card(J_{s})=n_{y}-q$ and for each set $S_{s}\subset\left\lbrace 1,\ldots,n_{y}\right\rbrace $ with $\card(S_{s})=n_{y}-2q$. Let Assumption \ref{c3a1} be satisfied, there exist at least one set $\bar{J}_{s}\subset\left\lbrace 1,\ldots,n_{y}\right\rbrace $ with $\card(\bar{J}_{s})=n_{y}-q$ such that $a_{y}^{J_{s}}(k)=0,\forall k\geq 0$. Then, the estimate given by the UIO for $\bar{J}_{s}$ is attack-free, and the estimates given by the UIOs for any $S_{s}\subset\bar{J}_{s}$ with $\card(S_{s})=n_{y}-2q$ which we denote as $\hat{x}_{S_{s}}$ are consistent with $\hat{x}_{J_{s}}$. This motivates the following estimation algorithm: for each set $J_{s}$ with $\card(J_{s})=n_{y}-q$, we define $\pi_{J_{s}}(k)$ as the largest deviation between $\hat{x}_{J_{s}}(k)$ and $\hat{x}_{S_{s}}(k)$ that is given by any $S_{s}\subset J_{s}$ with $\card(S_{s})=n_{y}-2q$, i.e.,
	\begin{equation}\label{c3es1}
		\pi_{J_{s}}(k)=\max_{S_{s}\subset J_{s}:\card(S_{s})=n_{y}-2q}|\hat{x}_{J_{s}}(k)-\hat{x}_{S_{s}}(k)|.
	\end{equation}
	For all $k\geq 0$,
	\begin{equation}\label{c33}
		\sigma_{s}(k)=\argmin_{J_{s}\subset\left\lbrace 1,\ldots,n_{y}\right\rbrace :\card(J_{s})=n_{y}-q}\pi_{J_{s}}(k),
	\end{equation}
	for all $k\geq 0$, the estimate given by $\sigma_{s}(k)$ is an attack-free estimate,
	\begin{equation}\label{c3es2}
		\hat{x}(k)=\hat{x}_{\sigma_{s}(k)}(k),
	\end{equation}
	where $\hat{x}_{\sigma_{s}(k)}(k)$ represents the estimates given by $\sigma_{s}(k)$. The above discussion is summarized as follows.
	\begin{theorem}\label{c3th1}
		Consider system (\ref{c3s1}), observer \eqref{c3o}, and the estimator (\ref{c3es1})-(\ref{c3es2}). Let Assumptions \ref{c3asump2}-\ref{c3a1} be satisfied and define the estimation error $e(k):=\hat{x}_{\sigma_s(k)}(k)-x(k)$; then, there exists a $\mathcal{KL}$-function $\bar{\beta}(\cdot,\cdot)$ satisfying:
		\begin{equation}
			\left\{\begin{split}
				|e(k)| &\leq\bar{\beta}(e_{0},k)\label{c3sa}\\
				e_{0} &:=\max_{ \tiny{\begin{array}{l} J_{s}:\card(J_{s})=n_{y}-q \\ S_{s}: \card(S_{s})=n_{y}-2q \end{array} } } \left\lbrace |e_{J_{s}}(0)|, |e_{S_{s}}(0)|\right\rbrace,
			\end{split}\right.
		\end{equation}
		for all $k\geq 0$.
	\end{theorem}
	\emph{\textbf{Proof:}}
	If Assumption \ref{c3a1} is satisfied, there exists at least one set $\bar{J}_{s}$ with $\card(\bar{J}_{s})=n_{y}-q$ such that $a_{y}^{\bar{J}_{s}}(k)=0$, $\forall k\geq 0$; then, there exists a $\mathcal{KL}$-function $\beta_{\bar{J}_{s}}(\cdot,\cdot)$ such that
	\begin{equation}\label{c36}
		|e_{\bar{J}_{s}}(k)|\leq \beta_{\bar{J}_{s}}(e_{0},k),
	\end{equation}
	for all $e_0\in\mathbb{R}^{n}$ and $k\geq0$. Also for any set $S_{s}\subset\bar{J}_{s}$ with $\card(S_{s})=n_{y}-2q$, we have $a_{y}^{S_{s}}(k)=0$, $\forall k\geq0$; hence, there exists a $\mathcal{KL}$-function $\beta_{S_{s}}(\cdot,\cdot)$ such that
	\begin{equation}\label{c361}
		|e_{S_{s}}(k)|\leq \beta_{S_{s}}(e_{0},k),
	\end{equation}
	for all $e_{0}\in\mathbb{R}^{n}$ and $k\geq 0$. From the definition of $\pi_{\bar{J}_{s}}$ in (\ref{c3es1}), we can write the following
	\begin{eqnarray}
		\begin{split}
			\pi_{\bar{J}_{s}}(k)=&\underset{S_{s}\subset\bar{J}_{s}}{\max}|\hat{x}_{\bar{J}_{s}}(k)-\hat{x}_{S_{s}}(k)|\\
			=&\underset{S_{s}\subset\bar{J}_{s}}{\max}|\hat{x}_{\bar{J}_{s}}(k)-x(k)+x(k)-\hat{x}_{S_{s}}(k)|\\
			\leq&  |e_{\bar{J}_{s}}(k)|+\underset{S_{s}\subset\bar{J}_{s}}{\max}|e_{S_{s}}(k)|
		\end{split}
	\end{eqnarray}
	for all $k\geq 0$. From (\ref{c36}) and (\ref{c361}), we obtain
	\begin{equation}\label{c3sv}
		\pi_{\bar{J}_{s}}(k)\leq 2\beta'_{\bar{J}_{s}}(e_{0},k),
	\end{equation}
	for all $e_{0}\in\mathbb{R}^{n}$ and $k\geq 0$, where \[\beta'_{\bar{J}_{s}}(e_{0},k):=\underset{S_{s}\subset\bar{J}_{s}}{\max}\left\lbrace \beta_{\bar{J}_{s}}(e_{0},k), \beta_{S_{s}}(e_{0},k)\right\rbrace, \] for all $k\geq 0$. From (\ref{c355}), $\pi_{\sigma_{s}(k)}(k)\leq\pi_{\bar{J}_{s}}(k)$. By Lemma \ref{c3l4}, at least one set $\bar{S}_{s}\subset\bar{J}_{s}$ with $\card(\bar{S}_{s})=n_{y}-2q$ exists such that $a_{y}^{\bar{S}_{s}}(k)=0$ for all $k\geq 0$, and there exists a $\mathcal{KL}$-function $\beta_{\bar{S}_{s}}(\cdot,\cdot)$ such that
	\begin{equation}\label{c3ss}
		|e_{\bar{S}_{s}}(k)|\leq \beta_{\bar{S}_{s}}(e_{0},k),
	\end{equation}
	for all $e_{0}\in\mathbb{R}^{n}$ and $k\geq0$. From (\ref{c3es1}), we have that \[
	\begin{split}
		\pi_{\sigma_{s}(k)}(k)=&\underset{S_{s}\subset\sigma_{s}(k)}{\max}|\hat{x}_{\sigma_{s}(k)}(k)-\hat{x}_{S_{s}}(k)|\\\geq&|\hat{x}_{\sigma_{s}(k)}(k)-\hat{x}_{\bar{S}_{s}}(k)|.
	\end{split}
	\] By the triangle inequality, we can write
	\begin{eqnarray}
		\begin{split}
			|e_{\sigma_{s}(k)}(k)|=&|\hat{x}_{\sigma_{s}(k)}(k)-x(k)|\\
			=&|\hat{x}_{\sigma_{s}(k)}(k)-\hat{x}_{\bar{S}_{s}}(k)+\hat{x}_{\bar{S}_{s}}(k)-x(k)|\\
			\leq&|\hat{x}_{\sigma_{s}(k)}(k)-\hat{x}_{\bar{S}_{s}}(k)|+|e_{\bar{S}_{s}}(k)|\\
			\leq&\pi_{\sigma_{s}(k)}(k)+|e_{\bar{S}_{s}}(k)|\\
			\leq&\pi_{\bar{J}_{s}}(k)+|e_{\bar{S}_{s}}(k)|
		\end{split}
	\end{eqnarray}
	for all $k\geq 0$. From (\ref{c3sv}) and (\ref{c3ss}), we have
	\begin{eqnarray}\label{c3sb}
		|e_{\sigma_{s}(k)}(k)|\leq \bar{\beta}(e_{0},k),
	\end{eqnarray}
	for all $e_{0}\in\mathbb{R}^{n}$ and $k\geq 0$, where \[\bar{\beta}(e_{0},k)=3\cdot \max\left\lbrace \beta_{\bar{S}_{s}}(e_{0},k),\beta'_{\bar{J}_{s}}(e_{0},k)\right\rbrace, \] for all $k\geq 0$. Inequality (\ref{c3sb}) is of the form (\ref{c3sa}) and the result follows.\hfill$\blacksquare$

	\subsection{Partial Unknown Input Observers}\label{c3partial}
	Let $B$ be partitioned as $B=\left[ b_{1},\hdots,b_{i},\hdots,b_{n_{u}}\right] $ with $b_{i}\in\mathbb{R}^{n\times 1}$ . Then, system (\ref{c3s1}) can be written as
	\begin{eqnarray}\label{c3s2}
		\begin{split}
			x^{+}=&f(x)+Bu+b_{W_{u}}a^{W_{u}},  \\
			y=&h(x)+a_{y},
		\end{split}
	\end{eqnarray}
	where we regard the vector of attacks $a^{W_{u}}$ as an unknown input to the dynamics. The columns of $b_{W_{u}}$ are all $b_{i}$ such that $i \in W_{u}$. Let $(q_{1},q_{2})$ be the largest integers such that a partial unknown input observer of the form
	\begin{equation}\label{c3o2}
		\hat{x}_{J_{us}}^{+}=f_{J_{us}}(\hat{x}_{J_{us}},u,y^{J_{s}},(y^{J_{s}})^{+}),
	\end{equation}
	exists for each $b_{J_{u}}$ and $J_{u}\subset \left\lbrace 1,\hdots,n_{u}\right\rbrace $ with $\card(J_{u})\leq 2q_{1}< n_{u}$ and each $y^{J_{s}}$ with $\card(J_{s})\geq n_{y}-2q_{2}>0$, where columns of $b_{J_{u}}$ are $b_{i}, i\in J_{u}$, i.e., an unknown input observer of the form \eqref{c3o2} exists for the following system:
	\begin{eqnarray}\label{c3s3}
		\begin{split}
			x^{+}=&f(x)+Bu+b_{J_{u}}a_{u}^{J_{u}},\\
			y^{J_{s}}=&h^{J_{s}}(x)+a_{y}^{J_{s}},
		\end{split}
	\end{eqnarray}
	with known input $u$ and unknown input $a_{u}^{J_{u}}$. UIOs of the form \eqref{c3o2} is referred to as \emph{partial} UIOs for the pair
	$(J_{u},J_{s})$.
	We assume the following.
	\begin{assumption}\label{c3asump1}
		At most $q_{1}$ actuators and $q_{2}$ sensors are under attack, i.e.,
		\begin{eqnarray}
			\card(W_{u})\leq q_{1}<\frac{n_{u}}{2},\\
			\card(W_{y})\leq q_{2}<\frac{n_{y}}{2},
		\end{eqnarray}
		where $q_{1}$ and $q_{2}$ denote the largest integers such that for any $J_{u}\subset\left\lbrace 1,\ldots,n_{u}\right\rbrace $ with $\card(J_{u})\leq 2q_{1}$ and $J_{s}\subset\left\lbrace 1,\ldots,n_{y}\right\rbrace $ with $\card(J_{s})\geq n_{y}-2q_{2}$, a partial UIO of the form \eqref{c3o2} exists for the pair $(J_{u},J_{s})$.
	\end{assumption}
	\begin{lemma}\label{c3lm2}
		If Assumption \ref{c3asump1} is satisfied, for each set of $q_{1}$ actuators, among all its supersets with $2q_{1}$ actuators, at least one set is a superset of $W_{u}$.
	\end{lemma}
	\begin{lemma}\label{c3lm4}
		If Assumption \ref{c3asump1} is satisfied, among each set of $n_{y}-q_{2}$ sensors, at least $n_{y}-2q_{2}>0$ of them are attack-free.
	\end{lemma}
	\emph{\textbf{Proof:}}
	Lemmas \ref{c3lm2} and \ref{c3lm4} follow trivially from Assumption \ref{c3asump1}.\hfill$\blacksquare$\\[1mm]
	We say that a UIO exists for each pair $(J_{u},J_{s})$ with $\card(J_{u})\leq 2q_{1}$ and $\card(J_{s})\geq n_{y}-2q_{2}$, if for $W_{u}\subseteq J_{u}$, $a_{y}^{J_{s}}(k)=0$, and $k\geq 0$, there exists a $\mathcal{KL}$-function $\beta_{J_{us}}(\cdot,\cdot)$ such that
	\begin{equation}
		|e_{J_{us}}(k)|\leq \beta_{J_{us}}(|e_{J_{us}}(0)|,k),
	\end{equation}
	where $e_{J_{us}}=\hat{x}_{J_{us}}-x$. We construct a partial UIO for each pair $(J_{u},J_{s})$ with $\card(J_{u})=q_{1}$ and $\card(J_{s})=n_{y}-q_{2}$ and for each pair $(S_{u},S_{s})$ with $\card(S_{u})=2q_{1}$ and $\card(S_{s})=n_{y}-2q_{2}$. Then, if Assumption \ref{c3asump1} is satisfied, there exists at least one set $\bar{J}_{u}$ with $\card(\bar{J}_{u})=q_{1}$ such that $W_{u}\subseteq\bar{J}_{u}$ and at least one set $\bar{J}_{s}$ with $\card(\bar{J}_{s})=n_{y}-q_{2}$ such that $a_{y}^{\bar{J}_{s}}(k)=0$, for all $k\geq 0$. Thus, the UIO for $(\bar{J}_{u},\bar{J}_{s})$ provides correct estimate, and the UIOs for any $(S_{u},S_{s})$ where $S_{u}\supset\bar{J}_{u}$ with $\card(S_{u})=2q_{1}$ and $S_{s}\subset \bar{J}_{s}$ with $\card(J_{s})=n_{y}-2q_{2}$ provide estimates (denotes as $\hat{x}_{S_{us}}$) that are consistent with $\hat{x}_{J_{us}}$. This motivates the following estimation strategy: for each $(J_{u},J_{s})$ with $\card(J_{u})=q_{1}$ and $\card(J_{s})=n_{y}-q_{2}$, we define $\pi_{J_{us}}(k)$ as the largest deviation between $\hat{x}_{J_{us}}(k)$ and $\hat{x}_{S_{us}}(k)$ that is given by any $(S_{u},S_{s})$ where $S_{u}\supset J_{u}$ with $\card(S_{u})=2q_{1}$ and  $S_{s}\subset J_{s}$ with $\card(S_{s})=n_{y}-2q_{2}$, i.e.,
	\begin{equation} \label{c354}
		\pi_{J_{us}}(k):=\max_{S_{u}\supset J_{u},S_{s}\subset J_{s}}|\hat{x}_{J_{us}}(k)-\hat{x}_{S_{us}}(k)|.
	\end{equation}
	for all $k\geq 0$, and
	\begin{equation}\label{c355}
		(\sigma_{u}(k),\sigma_{s}(k))=\underset{J_{u},J_{s}}{\argmin}\hspace{2mm} \pi_{J_{us}}(k);
	\end{equation}
	then, we say that the estimate given by ($\sigma_{u}(k),\sigma_{s}(k)$) is a correct estimate, i.e.,
	\begin{equation}\label{c356}
		\hat{x}(k)=\hat{x}_{\sigma_{us}(k)}(k),
	\end{equation}
	where $\hat{x}_{\sigma_{us}(k)}(k)$ denotes the estimate indexed by ($\sigma_{u}(k),\sigma_{s}(k)$). The above discussion is summarized in the following.
	
	\begin{theorem}\label{c3th2}
		Consider system (\ref{c3s1}), observer \eqref{c3o2}, and the estimator (\ref{c354})-(\ref{c356}). Let Assumption \ref{c3asump1} be satisfied and define the estimation error $e(k)=\hat{x}_{\sigma_{us}(k)}(k)-x(k)$; then, there exists a $\mathcal{KL}$-function $\bar{\beta}(\cdot,\cdot)$ satisfying:
		\begin{equation}
			\left\{\begin{split}
				|e(k)| &\leq\bar{\beta}(e_{0},k)\label{c362}\\
				e_{0} &:=\max_{ \tiny{\begin{array}{l} (J_{u},J_{s}) \\ (S_{u},S_{s})\end{array} } } \left\lbrace |e_{J_{us}}(0)|, |e_{S_{us}}(0)|\right\rbrace,
			\end{split}\right.
		\end{equation}
		for all $e_0\in\mathbb{R}^{n}$, $k\geq 0$.
	\end{theorem}
	\emph{\textbf{Proof:}}
	If Assumption \ref{c3asump1} is satisfied, there exist at least one set $\bar{J}_{u}$ with $\card(\bar{J})=q_{1}$ such that $\bar{J}_{u}\supset W_{u}$ and at least one set $\bar{J}_{s}$ with $\card(\bar{J}_{s})=n_{y}-q_{2}$ such that $a_{y}^{\bar{J}_{s}}(k)=0,\forall k\geq 0$, then, there exist a $\mathcal{KL}$-function $\beta_{\bar{J}_{us}}(\cdot,\cdot)$, such that
	\begin{equation}\label{c363}
		|e_{\bar{J}_{us}}(k)|\leq \beta_{\bar{J}_{us}}(e_{0},k),
	\end{equation}
	for all $e_{0}\in\mathbb{R}^{n}$ and $k\geq0$. Also for any set $S_{u}\supset\bar{J}_{u}$ with $\card(S_{u})=2q_{1}$ and $S_{s}\subset\bar{J}_{s}$ with $\card(S_{s})=n_{y}-2q_{2}$, we have $S_{u}\supset W_{u}$ and $a_{y}^{S_{s}}(k)=0$ $\forall k\geq0$, hence there exist a $\mathcal{KL}$-function $\beta_{S_{us}}(\cdot,\cdot)$, such that
	\begin{equation}\label{c361p}
		|e_{S_{us}}(k)|\leq \beta_{S_{us}}(e_{0},k),
	\end{equation}
	for all $e_{0}\in\mathbb{R}^{n}$ and $k\geq 0$. Recalling the definition of $\pi_{\bar{J}_{us}}$ from (\ref{c354}), we have that
	\begin{eqnarray}
		\begin{split}
			\pi_{\bar{J}_{us}}(k)&=\underset{S_{u}\supset\bar{J}_{u},S_{s}\subset\bar{J}_{s}}{\max}|\hat{x}_{\bar{J}_{us}}(k)-\hat{x}_{S_{us}}(k)|\\
			=&\underset{S_{u}\supset\bar{J}_{u},S_{s}\subset\bar{J}_{s}}{\max}|\hat{x}_{\bar{J}_{us}}(k)-x(k)+x(k)-\hat{x}_{S_{us}}(k)|\\
			\leq&  |e_{\bar{J}_{us}}(k)|+\underset{S_{u}\supset\bar{J}_{u},S_{s}\subset\bar{J}_{s}}{\max}|e_{S_{us}}(k)|
		\end{split}
	\end{eqnarray}
	for all $k\geq 0$. From (\ref{c363}) and (\ref{c361}), we obtain
	\begin{equation}\label{c366}
		\pi_{\bar{J}_{us}}(k)\leq 2\beta'_{\bar{J}_{us}}(e_{0},k),
	\end{equation}
	for all $e_{0}\in\mathbb{R}^{n}$ and $k\geq 0$, where \[\beta'_{\bar{J}_{us}}(e_{0},k):=\underset{S_{u}\supset\bar{J}_{u},S_{s}\subset\bar{J}_{s}}{\max}\left\lbrace \beta_{\bar{J}_{us}}(e_{0},k), \beta_{S_{us}}(e_{0},k)\right\rbrace,\] 
	for all $k\geq 0$. Recall from (\ref{c355}) that $\pi_{\sigma_{us}(k)}(k)\leq\pi_{\bar{J}_{us}}(k)$. From Lemmas \ref{c3lm2}, \ref{c3lm4}, we know that there exist at least one set $\bar{S}_{u}\supset\sigma(k)$ with $\card(\bar{S}_{u})=2q_{1}$ and at least one set $\bar{S}_{s}\subset\bar{J}_{s}$ with $\card(\bar{S}_{s})=n_{y}-2q_{2}$ such that $\bar{S}_{u}\supset W_{u}$ and $a_{y}^{\bar{S}_{s}}(k)=0$ for all $k\geq 0$, and there exist a class $\mathcal{KL}$-function $\beta_{\bar{S}_{us}}(\cdot,\cdot)$, such that 
	\begin{equation}\label{c367}
		|e_{\bar{S}_{us}}(k)|\leq \beta_{\bar{S}_{us}}(e_{0},k),
	\end{equation}
	for all $e_{0}\in\mathbb{R}^{n}$ and $k\geq0$. From (\ref{c354}), there is a fact that \[
	\begin{split}
		\pi_{\sigma_{us}(k)}(k)=&\underset{S_{u}\supset \sigma_{u}(k),S_{s}\subset\sigma_{s}(k)}{\max}|\hat{x}_{\sigma_{us}(k)}(k)-\hat{x}_{S_{us}}(k)|\\\geq&|\hat{x}_{\sigma_{us}(k)}(k)-\hat{x}_{\bar{S}_{us}}(k)|.
	\end{split}
	\] From the triangle inequality we have that 
	\begin{eqnarray}
		\begin{split}
			|e_{\sigma_{us}(k)}(k)|=&|\hat{x}_{\sigma_{us}(k)}(k)-x(k)|\\
			=&|\hat{x}_{\sigma_{us}(k)}(k)-\hat{x}_{\bar{S}_{us}}(k)+\hat{x}_{\bar{S}_{us}}(k)-x(k)|\\
			\leq&|\hat{x}_{\sigma_{us}(k)}(k)-\hat{x}_{\bar{S}_{us}}(k)|+|e_{\bar{S}_{us}}(k)|\\
			\leq&\pi_{\sigma_{us}(k)}(k)+|e_{\bar{S}_{us}}(k)|\\
			\leq&\pi_{\bar{J}_{us}}(k)+|e_{\bar{S}_{us}}(k)|
		\end{split}
	\end{eqnarray}
	for all $k\geq 0$. From (\ref{c366}) and (\ref{c367}), we have 
	\begin{eqnarray}\label{c370}
		|e_{\sigma_{us}(k)}(k)|\leq \bar{\beta}(e_{0},k),
	\end{eqnarray}
	for all $e_{0}\in\mathbb{R}^{n}$ and $k\geq 0$, where \[\bar{\beta}(e_{0},k)=3\cdot \max\left\lbrace \beta_{\bar{S}_{us}}(e_{0},k),\beta'_{\bar{J}_{us}}(e_{0},k)\right\rbrace .\] (\ref{c370}) is of the form (\ref{c362}) and the result follows.\hfill$\blacksquare$
	
	\subsection{An Application Example}
	Consider the nonlinear system:
	\begin{equation}
		\begin{split}
			x^{+}=&Ax+f(x)+B(u+a_{u}),\\
			y=&Cx+a_{y},
		\end{split}
	\end{equation}
	with matrix $C\in\mathbb{R}^{n_{y}\times n}$ and nonlinear function $f:\mathbb{R}^{n}\rightarrow\mathbb{R}^{n}$ satisfying the following Lipschitz condition:
	\begin{equation}
		\left| f(x_{1})-f(x_{2})\right|\leq\gamma|x_{1}-x_{2}|,\forall x_{1},x_{2}\in\mathbb{R}^{n},
	\end{equation}
	where $\gamma>0$ denotes the Lipschitz constant. Consider a complete UIO of the form:
	\begin{equation}
		\begin{split}
			\hat{x}_{J_{s}}^{+}=&\bar{A}_{J_{s}}\hat{x}_{J_{s}}+\bar{B}_{J_{s}}u+\bar{f}_{J_{s}}(\hat{x}_{J_{s}})+K_{J_{s}}(y^{J_{s}}-C^{J_{s}}\hat{x}_{J_{s}})\\
			&+\bar{B}_{J_{s}}(y^{J_{s}})^+,
		\end{split}
	\end{equation}
	where $K_{J_{s}}\in\mathbb{R}^{n\times\card(J_{s})}$ is the observer gain. Let $H_{J_{s}}:=(C^{J_s}B)_{left}^{-1}$, $\bar{G}_{J_{s}}:=I-BH_{J_{s}}C^{J_{s}}$, $\bar{A}_{J_{s}}:=\bar{G}_{J_{s}}A$, $\bar{B}_{J_{s}}=\bar{G}_{J_{s}}B$, and $\bar{f}_{J_{s}}(\cdot)=\bar{G}_{J_s}f(\cdot)$. If for all $J_{s}\subset\left\lbrace 1,\ldots,n_{y}\right\rbrace $ with $\card(J_{s})\geq n_{y}-2q$, it is satisfied that $\rank(C^{J_{s}}B)=n_{u}$; then, complete UIOs can be designed using the tools given in \cite{Witczak2007} for all $y^{J_{s}}$ with $\card(J_{s})\geq n_{y}-2q$. Using the estimation strategy (\ref{c3es1})-(\ref{c3es2}) and Theorem \ref{c3th1}, we can conclude that (\ref{c3sa}) is satisfied for all $e_0 \in \mathbb{R}^{n}$ and $k\geq 0$. If $n_{y}-2<n_{u}$; then, complete UIOs cannot be designed for any $y^{J_{s}}$ with $\card(J_{s})= n_{y}-2$ using the design methods given in \cite{Witczak2007}. Then, in that case, consider partial UIOs of the form:
	\begin{equation}
		\begin{split}
			\hat{x}_{J_{us}}^{+}=&\bar{A}_{J_{us}}\hat{x}_{J_{us}}+\bar{B}_{J_{us}}u+\bar{f}_{J_{us}}(\hat{x}_{J_{us}})\\
			&+K_{J_{us}}(y^{J_{s}}-C^{J_{s}}\hat{x}_{J_{us}})+\bar{b}_{J_{us}}(y^{J_{s}})^+.
		\end{split} 
	\end{equation}
	Let $H_{J_{us}}:=(C^{J_{s}}b_{J_{u}})^{-1}_{left}$, $\bar{G}_{J_{us}}:=I-b_{J_u}H_{J_{us}}C^{J_{s}}$, $\bar{b}_{J_{us}}:=b_{J_{u}}H_{J_{us}}$, $\bar{A}_{J_{us}}:=\bar{G}_{J_{us}}A$, $\bar{B}_{J_{us}}:=\bar{G}_{J_{us}}B$, and $\bar{f}_{J_{us}}(\cdot):=\bar{G}_{J_{us}}f(\cdot)$. If for all $J_{u}\subset\left\lbrace 1,\ldots,n_{u}\right\rbrace $, $\card(J_{u})\leq 2q_{1}$, and $J_{s}\subset\left\lbrace 1,\ldots,n_{y}\right\rbrace $, $\card(J_{s})\geq n_{y}-2q_{2}$, it is satisfied that $\rank(C^{J_{s}}b_{J_{u}})=\rank(b_{J_u})=\card(J_{u})$; then, partial UIOs can be designed using the method given in \cite{Witczak2007}, for all $(J_{u},J_{s})$ with $\card(J_{u})\leq 2q_{1}$, $\card(J_{s})\geq n_{y}-2q_{2}$. Under Assumption \ref{c3asump1}, using the estimation strategy (\ref{c354})-(\ref{c356}) and Theorem \ref{c3th2}, we can conclude that (\ref{c362}) is satisfied for all $e_0\in\mathbb{R}^{n}$ and $k\geq 0$.
	
\begin{figure}[t]\centering
		\includegraphics[width=0.45\textwidth]{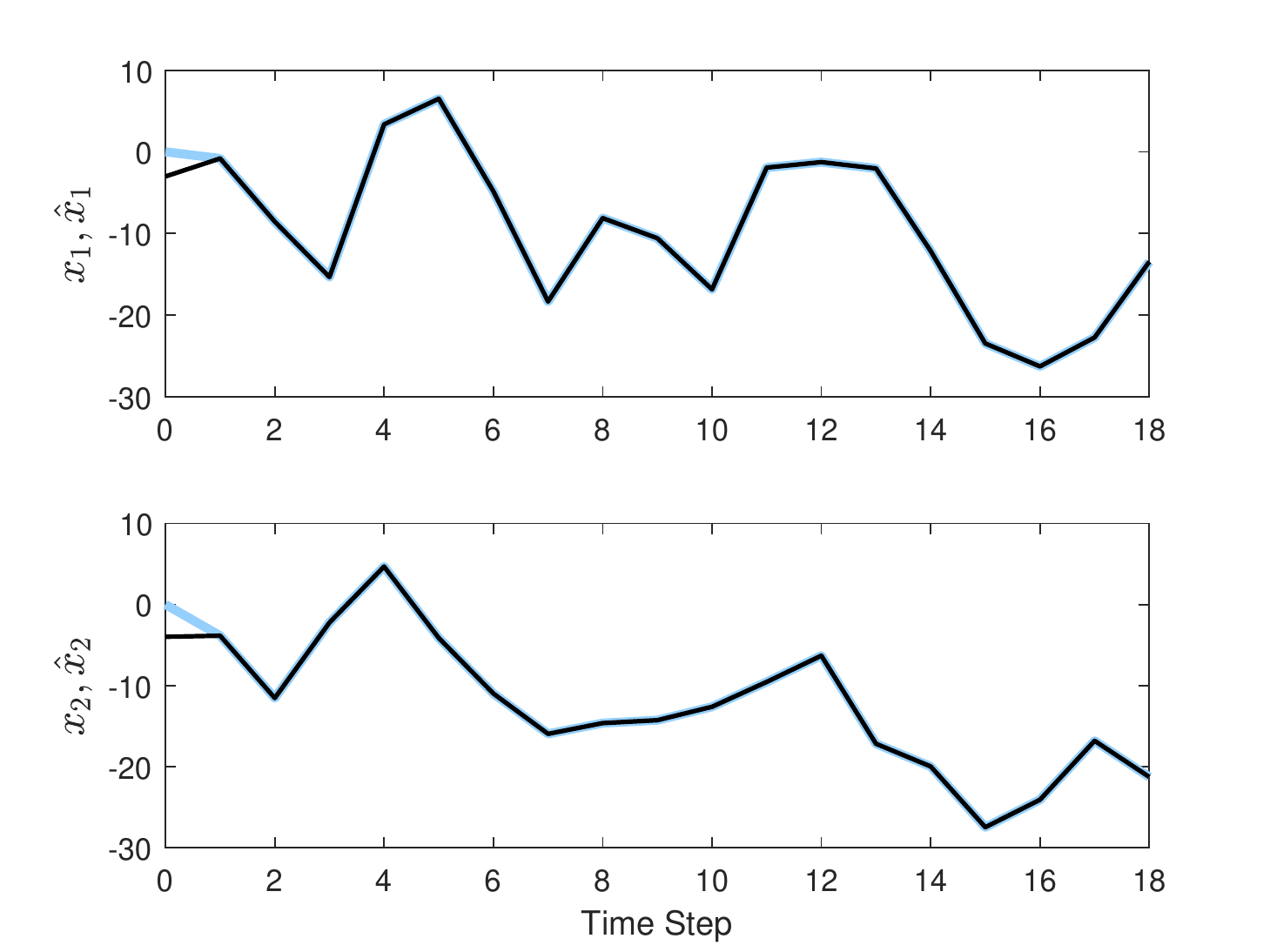}
		\caption{The estimation $\hat{x}$ converges to the true states $x$ when $a_{u1},a_{u2},a_{y2}\sim\mathcal{U}(-10,10)$. Legend: $\hat{x}$ (grey), true states (black)}
		\label{c3f2}
		\centering
	\end{figure}

	\textbf{Example 1.} Consider the nonlinear system under sensor and actuator attacks:
	\begin{equation}\label{c3e1}
		\begin{split}
			x^{+}=&\left[ \begin{matrix}
				0.2&0.5\\0.2&0.7
			\end{matrix}\right]x +\left[ \begin{matrix}
				0.5\sin{x_{1}}\\
				0.5\sin{x_{2}}
			\end{matrix}\right] +\left[ \begin{matrix}
				1&0\\0&1
			\end{matrix}\right] (u+a_{u}),\\
			y=&\left[ \begin{matrix}
				1&1&3&4\\
				1&1&2&1
			\end{matrix}\right]^{\top}x+a_{y}.
		\end{split}
	\end{equation}
	Using the method given in \cite{Witczak2007}, a complete UIO can be designed for each $y^{J_{s}}$ with $\card(J_{s})\geq 2$. Therefore, we have $q=1$. We let $W_{u}=\left\lbrace 1,2\right\rbrace $, which means both actuators are under attack, and $W_{y}=\left\lbrace 2\right\rbrace $, which means the $2$-nd sensor is compromised. We let $(u_{1},u_{2})\sim\mathcal{U}(-5,5)$, $(a_{u1},a_{u2},a_{y2})\sim\mathcal{U}(-10,10)$. Then, we design a complete UIO for each $J_{s}\subset\left\lbrace 1,2,3,4\right\rbrace $ with $\card(J_{s})=3$ and each $S_{s}\subset\left\lbrace 1,2,3,4\right\rbrace $ with $\card(S_{s})=2$. Therefore, $\binom{4}{3}+\binom{4}{2}=10$ complete UIOs are designed in total, which are all initialized at $\left[ 0,0\right] ^{\top}$. For all $k\geq 0$, $(\ref{c3es1})-(\ref{c3es2})$ is used to construct $\hat{x}(k)$. The performance of the estimator is shown in Figures \ref{c3f2}-\ref{c3f2}.
	\begin{figure}[t]\centering
		\includegraphics[width=0.45\textwidth]{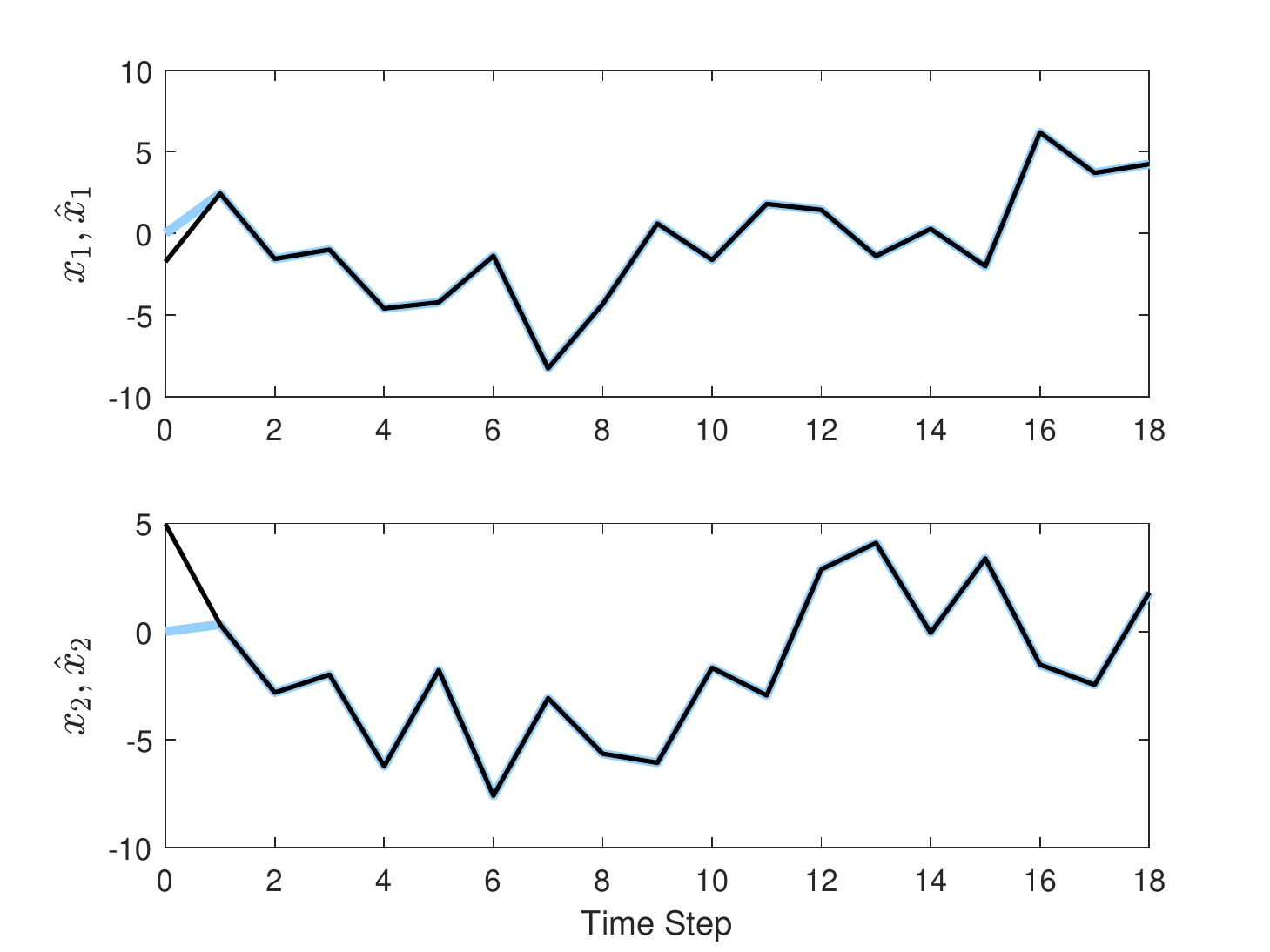}
		\caption{The estimation $\hat{x}$ converges to the true states $x$ when $a_{u1},a_{u2},a_{y2}\sim\mathcal{U}(-1,1)$. Legend: $\hat{x}$ (grey), true states (black)}
		\label{c3f1}
		\centering
	\end{figure}
	
\begin{figure}[t]\centering
		\includegraphics[width=0.45\textwidth]{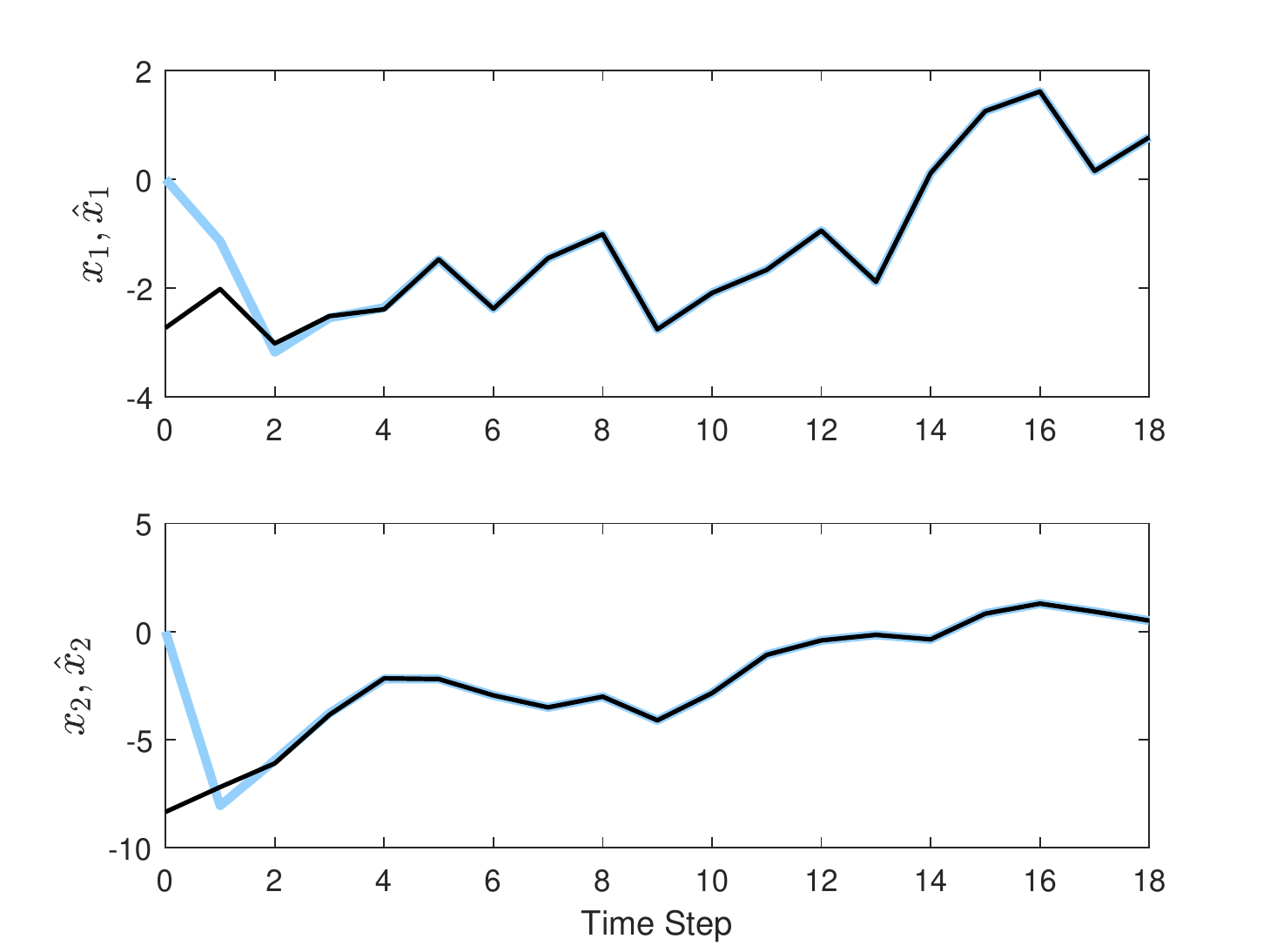}
		\caption{The estimation $\hat{x}$ converges to the true states $x$ when $a_{u1},a_{y2}\sim\mathcal{U}(-1,1)$. Legend: $\hat{x}$ (grey), true states (black)}
		\label{c3f3}
		\centering
	\end{figure}

\begin{figure}[t]\centering
		\includegraphics[width=0.45\textwidth]{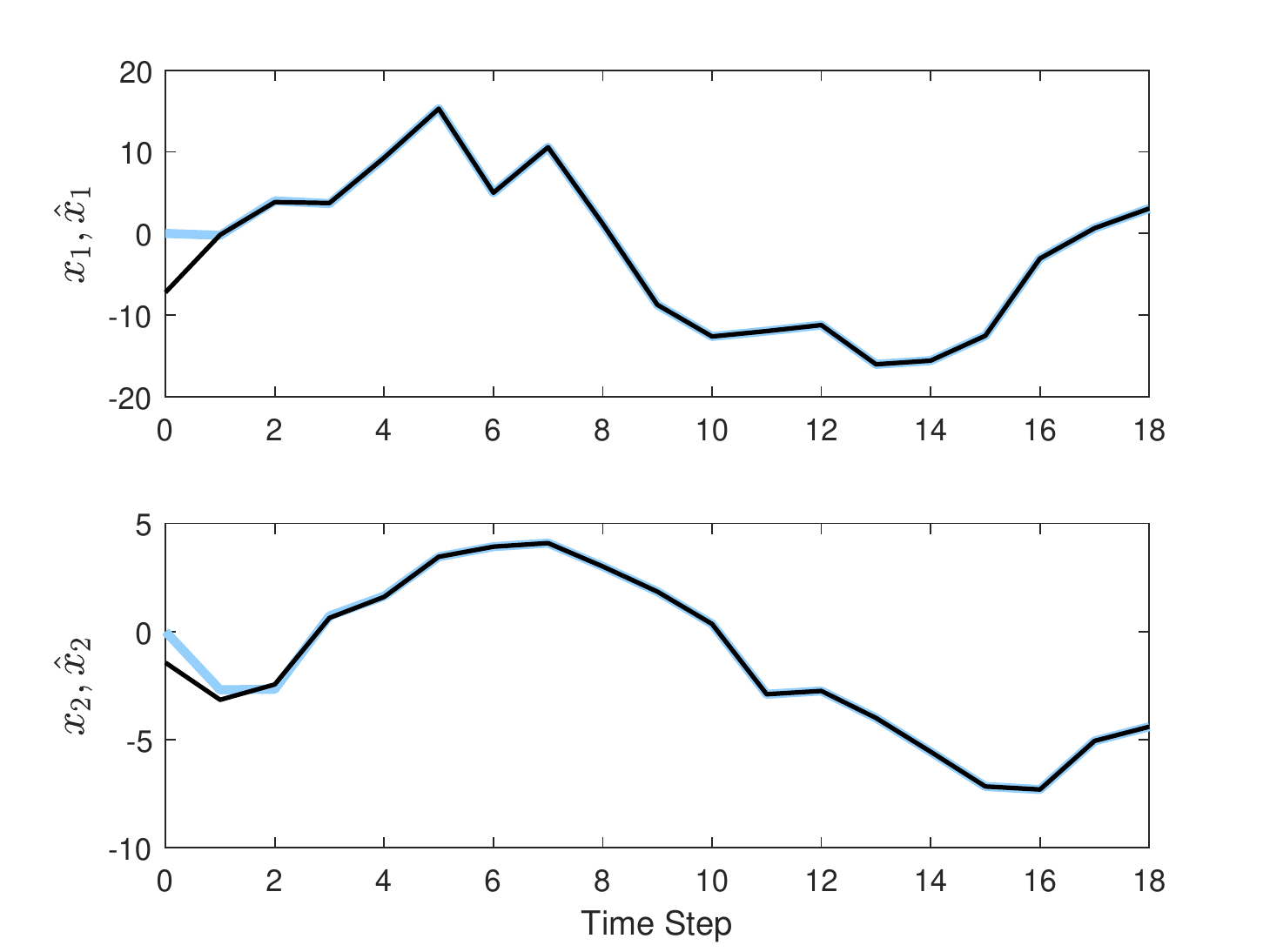}
		\caption{The estimation $\hat{x}$ converges to the true states $x$ when $a_{u1},a_{y2}\sim\mathcal{U}(-10,10)$. Legend: $\hat{x}$ (grey), true states (black)}
		\label{c3f4}
		\centering
	\end{figure}

	\textbf{Example 2.}\label{c3ex2}
	Consider the nonlinear system:
	\begin{eqnarray}
		\begin{split}
			x^{+}=&\left[ \begin{matrix}\label{c3e2}
				0.5&0&0.1\\
				0.2&0.7&0\\
				1&0&0.3
			\end{matrix}\right] x+\left[ \begin{matrix}
				0.5\sin{x_{1}}\\
				0.5\sin{x_{2}}\\
				0.5\sin{x_{3}}
			\end{matrix}\right]
			\\
			&+ \left[ \begin{matrix}
				1&0&1\\
				1&1&0\\
				0&1&1
			\end{matrix}\right] (u+a_{u}),\\
			y=&\left[ \begin{matrix}
				1&1&3&4\\3&1&2&1\\1&1&1&1
			\end{matrix}\right]^{\top} x+a_{y}.
		\end{split}
	\end{eqnarray}
	We have $n_{y}=4$ and $n_{u}=3$; then, $n_{y}-2<n_{u}$ and it can be verified that complete UIOs cannot be designed for any $y^{J_{s}}$ with $\card(J_{s})=2$ using the design methods given in \cite{Witczak2007}. Instead, partial UIOs can be designed for each pair $(J_{u}, J_{s})$ with $\card(J_{u})\leq 2$ and $\card(J_{s})\geq 2$. We let $q_{1}=q_{2}=1$, $(u_{1},u_{2},u_{3})\sim\mathcal{U}(-1,1)$, $W_{u}=\left\lbrace 1\right\rbrace $, $W_{y}=\left\lbrace 2\right\rbrace $, $(a_{u3},a_{y2})\sim\mathcal{U}(-10,10)$. We construct a partial UIO for each set pair $(J_{u},J_{s})$ with $\card(J_{u})=1,\card(J_{s})=3$ and each set pair $(S_{u},S_{s})$ with $\card(S_{u})=2,\card(S_{s})=2$. Therefore, totally $\binom{3}{1}\times\binom{4}{3}+\binom{3}{2}\times\binom{4}{2}=30$ partial UIOs are constructed and we initialize them by letting $\hat{x}(0)=[0,0]^{\top}$. For all $k\geq 0$, (\ref{c354})-(\ref{c356}) is used to construct $\hat{x}(k)$. We depict the performance of the estimator in Figures \ref{c3f3}-\ref{c3f4}.
	\section{Isolation of Attacks}\label{isolation}
	The estimate $\hat{x}(k)$ of $x(k)$, provided by the estimator in Section \ref{c3complete} or the one in Section \ref{c3partial}, can be used combined with the system dynamics \eqref{c3s1}, and the known inputs to asymptotically reconstruct the attack signals. Note that $e = \hat{x} - x \Rightarrow x = \hat{x} - e \Rightarrow x^{+} = \hat{x}^{+} - e^{+}$. Then, we reformulate the system dynamics \eqref{c3s1} in terms of $e$ and $\hat{x}$ as follows:
	\begin{equation}\label{c3attack_estimation_1}
		\left\{
		\begin{split}
			\hat{x}^{+} &= e^{+} + A(\hat{x}-e)+f(\hat{x} - e) + B(u+a_{u}),\\
			&\hspace{25mm}\Downarrow\\
			a_{u} &= B_{left}^{-1}(\hat{x}^{+}-A\hat{x}-f(\hat{x}-e))-u-B_{left}^{-1}(e^{+}+Ae), \\
		\end{split}
		\right.
	\end{equation}
	where, because $B$ has full column rank, $B_{left}^{-1}$ denotes the Moore-Penrose pseudoinverse of $B$. Similarly, we have
	\begin{equation}\label{c3attack_estimation_2}
		\left\{
		\begin{split}
			y&= Cx+a_{y}=C\hat{x}-Ce+a_{y},\\
			&\hspace{25mm}\Downarrow\\
			a_{y} &= y-C\hat{x}+Ce. \\
		\end{split}
		\right.
	\end{equation}
	We first consider the complete multi-observer in Section \ref{c3complete}. Suppose the dynamics of the estimation error characterized by \eqref{c3es1}-\eqref{c3es2} is as:
	\begin{equation}\label{c3he}
		e^{+}=f_{1}(e,x,a_{y},a_{y}^{+},a_{u}),
	\end{equation}
	where $f_{1}:\mathbb{R}^n \times \mathbb{R}^n \times \mathbb{R}^{n_{y}}\times \mathbb{R}^{n_{y}}\times\mathbb{R}^{n_{u}} \rightarrow \mathbb{R}^n$ denotes some nonlinear function. In Theorem \ref{c3th1}, we have proved that $e$ is asymptotically stable. Therefore, the terms that depend on $e$ and $e^+$ in the expression for $a_{u}$ and $a_{y}$ in \eqref{c3attack_estimation_1} and \eqref{c3attack_estimation_2}, respectively, vanish asymptotically and hence, the following formulas:
	\begin{equation}\label{c3ru}
		\hat{a}_{u}(k) = B_{left}^{-1}(\hat{x}(k) - A\hat{x}(k-1)-f(\hat{x}(k-1)))-u(k-1),
	\end{equation}
	and
	\begin{equation}\label{c3ry}
		\hat{a}_{y}(k)=y(k)-C\hat{x}(k),
	\end{equation}
	provide an asymptotically reconstruction of the attack signals $a_{u}(k-1)$ and $a_{y}(k)$, i.e.,
	\begin{equation}
		\begin{split}
			\lim_{k\to\infty}(\hat{a}_{u}(k)-a_{u}(k-1))=0,
		\end{split}
	\end{equation}
	and
	\begin{equation}
		\lim_{k\to\infty}(\hat{a}_{y}(k)-a_{y}(k))=0.
	\end{equation}
	Then, by simply checking the sparsity pattern of $\hat{a}_{u}(k)$ and $\hat{a}_{y}(k)$, we can isolate attacks for sufficiently large $k$, i.e.,
	\begin{equation}\label{c3wu}
		\hat{W}_{u}(k)=\supp(\hat{a}_{u}(k)),
	\end{equation}
	and
	\begin{equation}\label{c3wy}
		\hat{W}_{y}(k)=\supp(\hat{a}_{y}(k)),
	\end{equation}
	where $\hat{W}_{u}(k)$ represents the set of attacked actuators we isolate, and $\hat{W}_{y}(k)$ represents the set of attacked sensors we isolate. Note that since $a_{u}$ is estimated from $\hat{x}^{+}$ and $e^{+}$, there is always at least, one-step delay for actuator attacks isolation.
	
	Next, consider the partial multi-observer estimator given in Section \ref{c3partial}. Similarly, we also write the attack vector $a_{u}$ and $a_{y}$ as \eqref{c3attack_estimation_1} and \eqref{c3attack_estimation_2}, and use some nonlinear difference equation to describe the estimation error dynamics characterized by the estimator (\ref{c354})-(\ref{c356}), which is given as follows:
	\begin{equation}\label{c3c4}
		e^{+}=f_{2}(e,x,a_{y},a_{y}^{+},a_{u}),
	\end{equation}
	where $f_{2}:\mathbb{R}^n \times \mathbb{R}^n \times \mathbb{R}^{n_{y}}\times \mathbb{R}^{n_{y}}\times\mathbb{R}^{n_{u}} \rightarrow \mathbb{R}^n$ is a nonlinear function. In Theorem \ref{c3th2}, we have proved that $e$ is asymptotically stable. Hence, the estimated attack signals given by \eqref{c3ru} and \eqref{c3ry} reconstruct the attack signals asymptotically. By checking the sparsity pattern of $\hat{a}_{u}(k)$ and $\hat{a}_{y}(k)$, we can effectively pinpoint attacked actuators and sensors using \eqref{c3wu} and \eqref{c3wy}.\\[1mm]
	
\begin{figure}[t]\centering
		\includegraphics[width=0.45\textwidth]{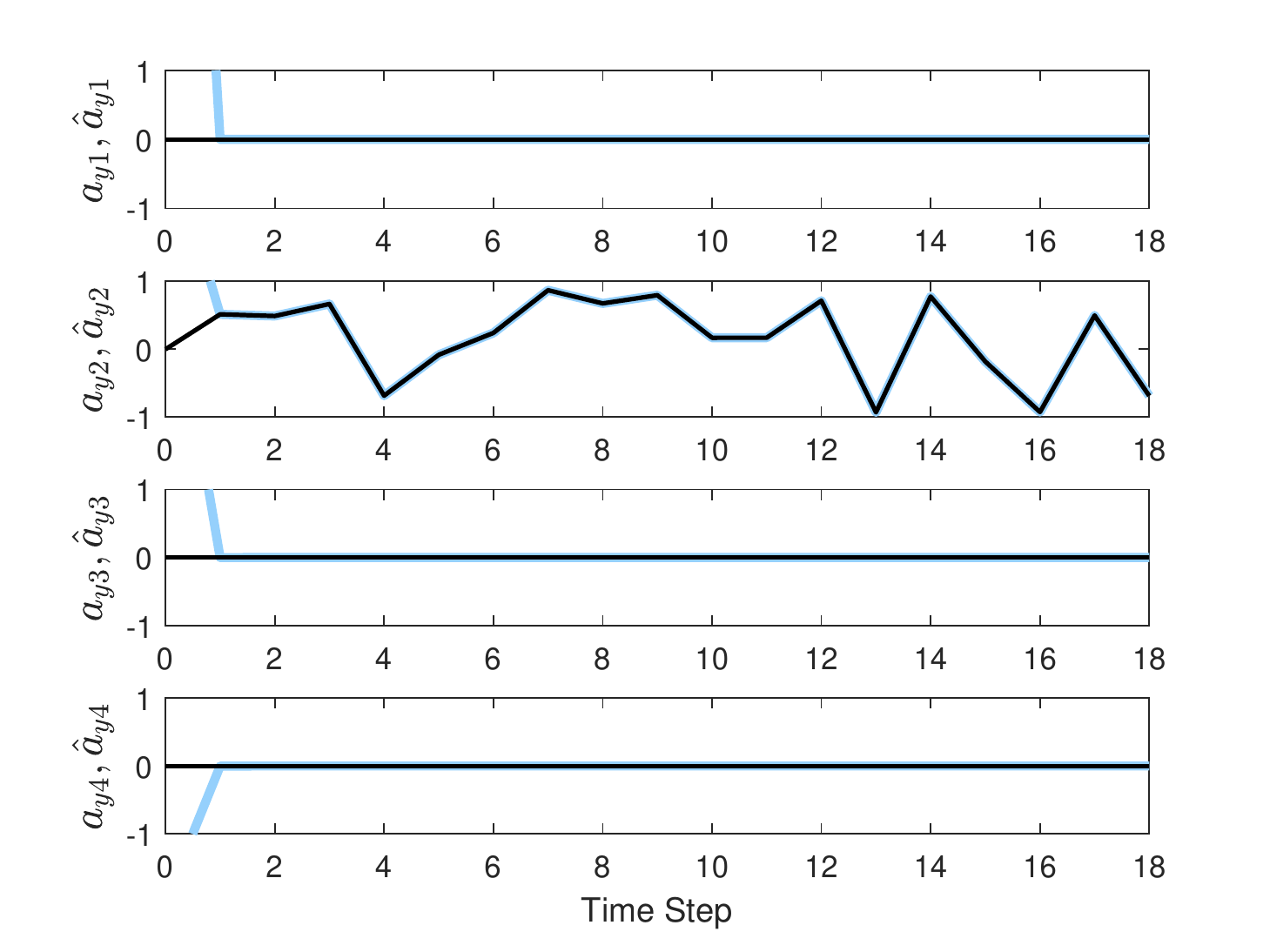}
		\caption{Estimate of $a_{y}$ when $a_{u1},a_{u2},a_{y2}\sim\mathcal{U}(-1,1)$. }
		\label{c3fig:3e}
		\centering
	\end{figure}

\begin{figure}[t]\centering
		\includegraphics[width=0.45\textwidth]{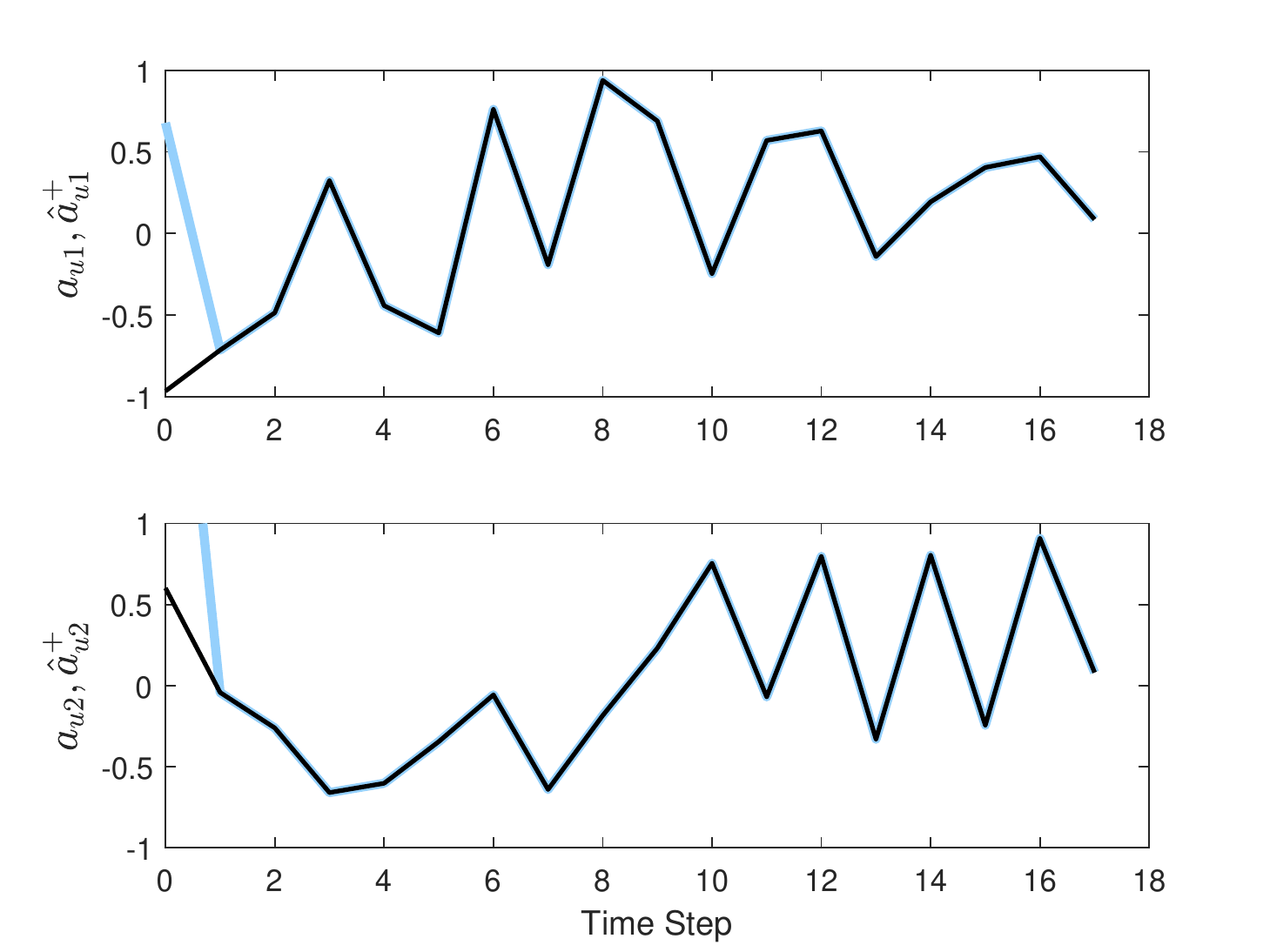}
		\caption{Estimate of $a_{u}$ when $a_{u1},a_{u2},a_{y2}\sim\mathcal{U}(-1,1)$. }
		\label{c3fig:4e}
		\centering
	\end{figure}

\begin{figure}[t]\centering
		\includegraphics[width=0.45\textwidth]{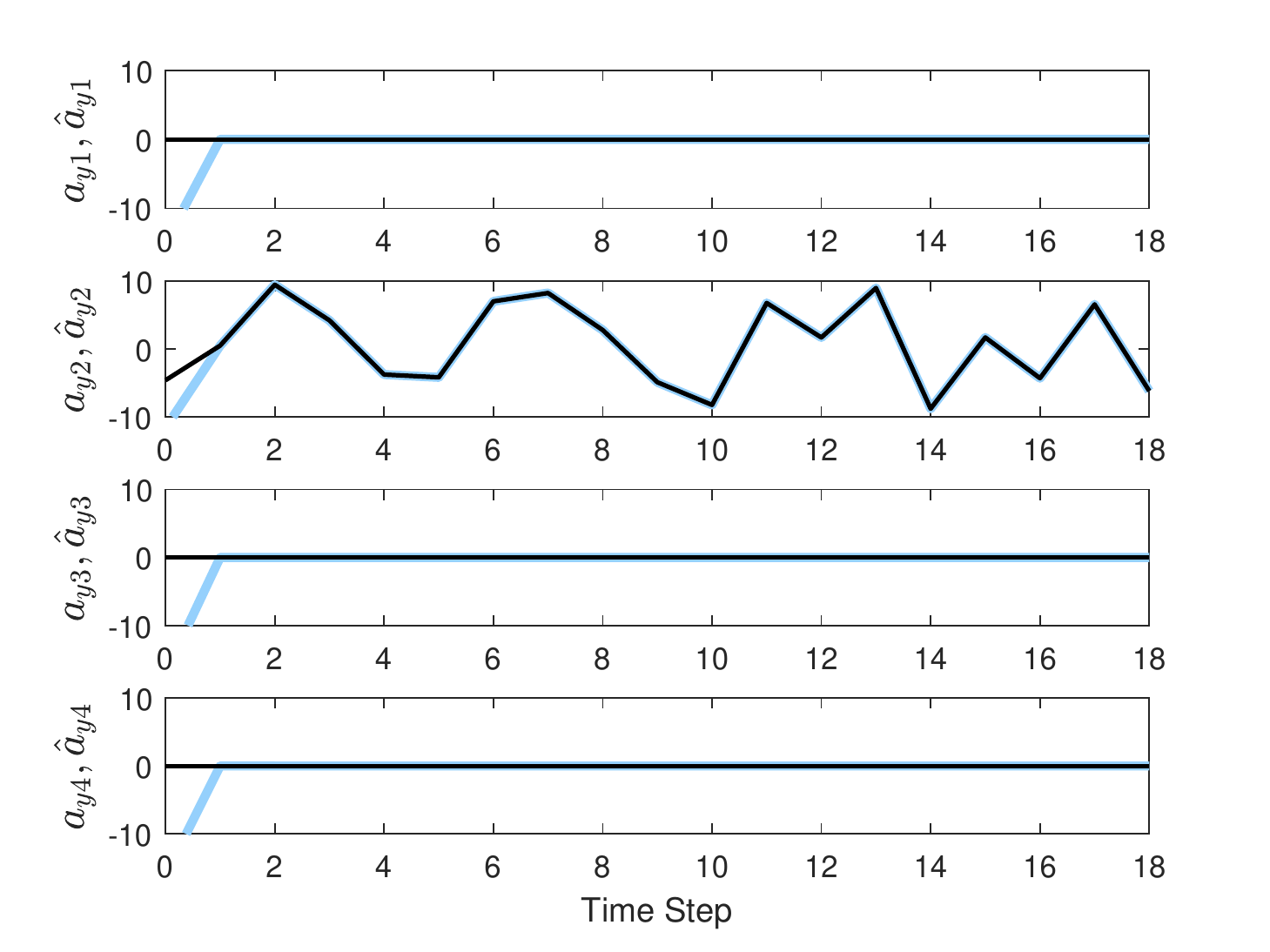}
		\caption{Estimate of $a_{y}$ when $a_{u1},a_{u2},a_{y2}\sim\mathcal{U}(-10,10)$. }
		\label{c3fig:5e}
		\centering
	\end{figure}

\begin{figure}[t]\centering
		\includegraphics[width=0.45\textwidth]{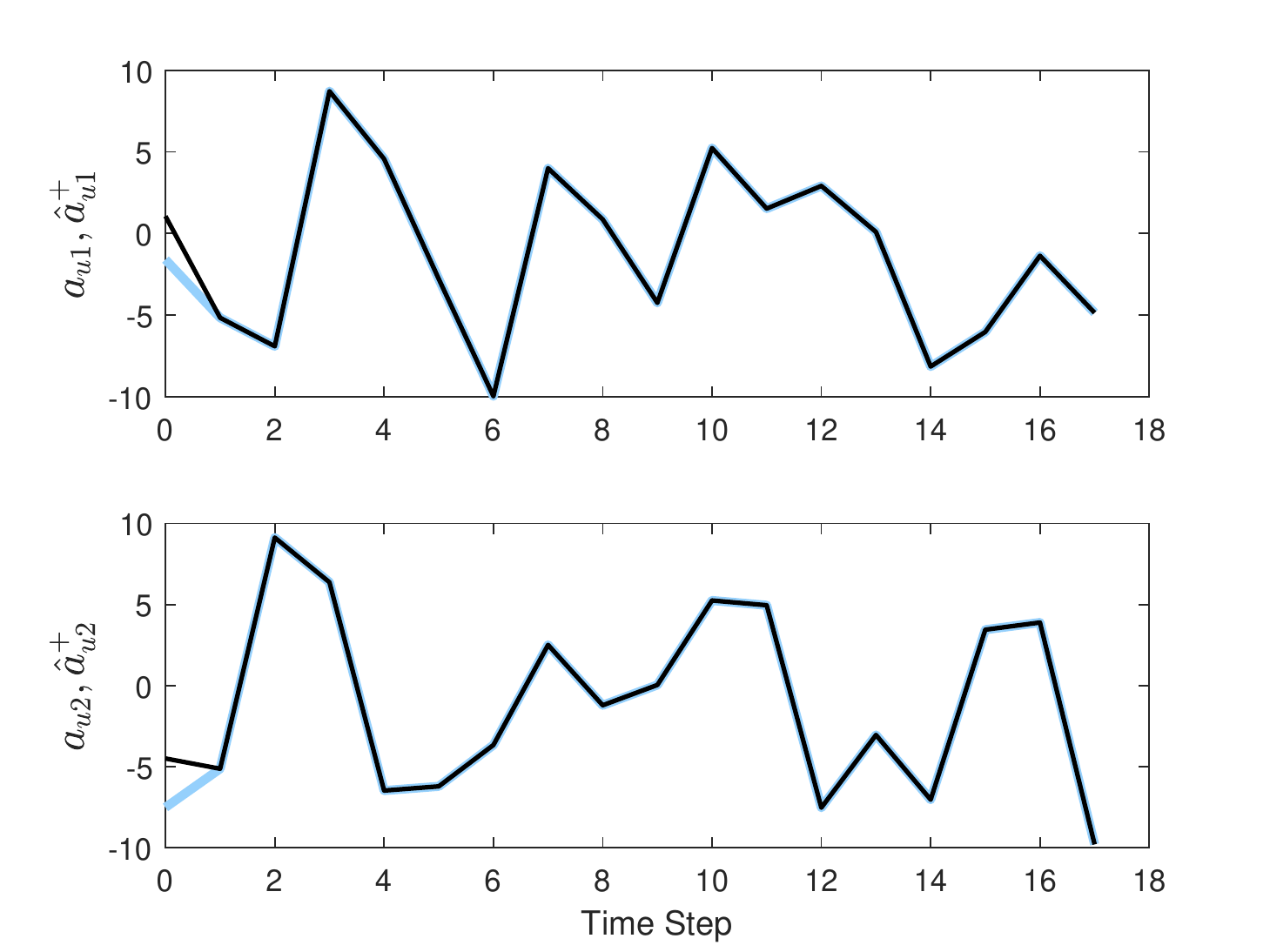}
		\caption{Estimate of $a_{u}$ when $a_{u1},a_{u2},a_{y2}\sim\mathcal{U}(-10,10)$. }
		\label{c3fig:6e}
		\centering
	\end{figure}

	\textbf{Example 3}
	We consider model (\ref{c3e1}) in Example 1. We let $q=1$, $W_{u}=\left\lbrace 1,2\right\rbrace $, $W_{y}=\left\lbrace 2\right\rbrace$, $(u_{1},u_{2})\sim\mathcal{U}(-5,5)$, $(a_{u1},a_{u2},a_{y2})\sim\mathcal{U}(-10,10)$, and $(x_{1}(0),x_{2}(0))\sim\mathcal{N}(0,1)$. We run $\binom{4}{3}+\binom{4}{2}=10$ complete UIOs initialized at $\hat{x}(0)=\left[ 0,0\right] ^{\top}$. We reconstruct $a_{y}$ and $a_{u}$ from (\ref{c3ry}) and (\ref{c3ru}) in $19$ time-steps. The performance of the attack estimation is shown in Figures \ref{c3fig:3e}-\ref{c3fig:6e}. By checking the sparsity, actuators $1$ and $2$ and sensor $2$ can be isolated as the attacked ones.
	
\begin{figure}[t]\centering
		\includegraphics[width=0.45\textwidth]{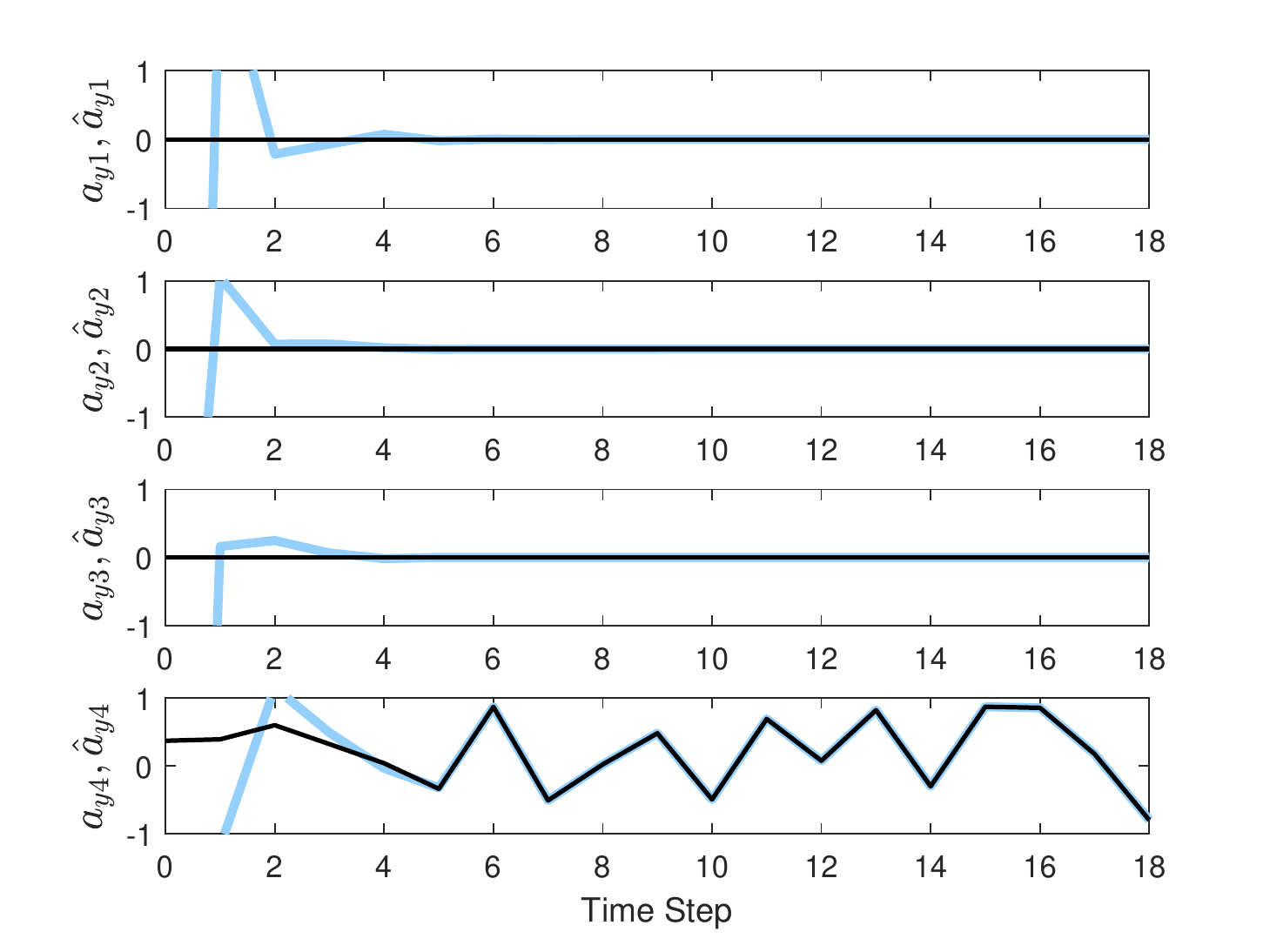}
		\caption{Estimate of $a_{y}$ when $a_{u3},a_{y2}\sim\mathcal{U}(-1,1)$. }
		\label{c331}
		\centering
	\end{figure}

\begin{figure}[t]\centering
		\includegraphics[width=0.45\textwidth]{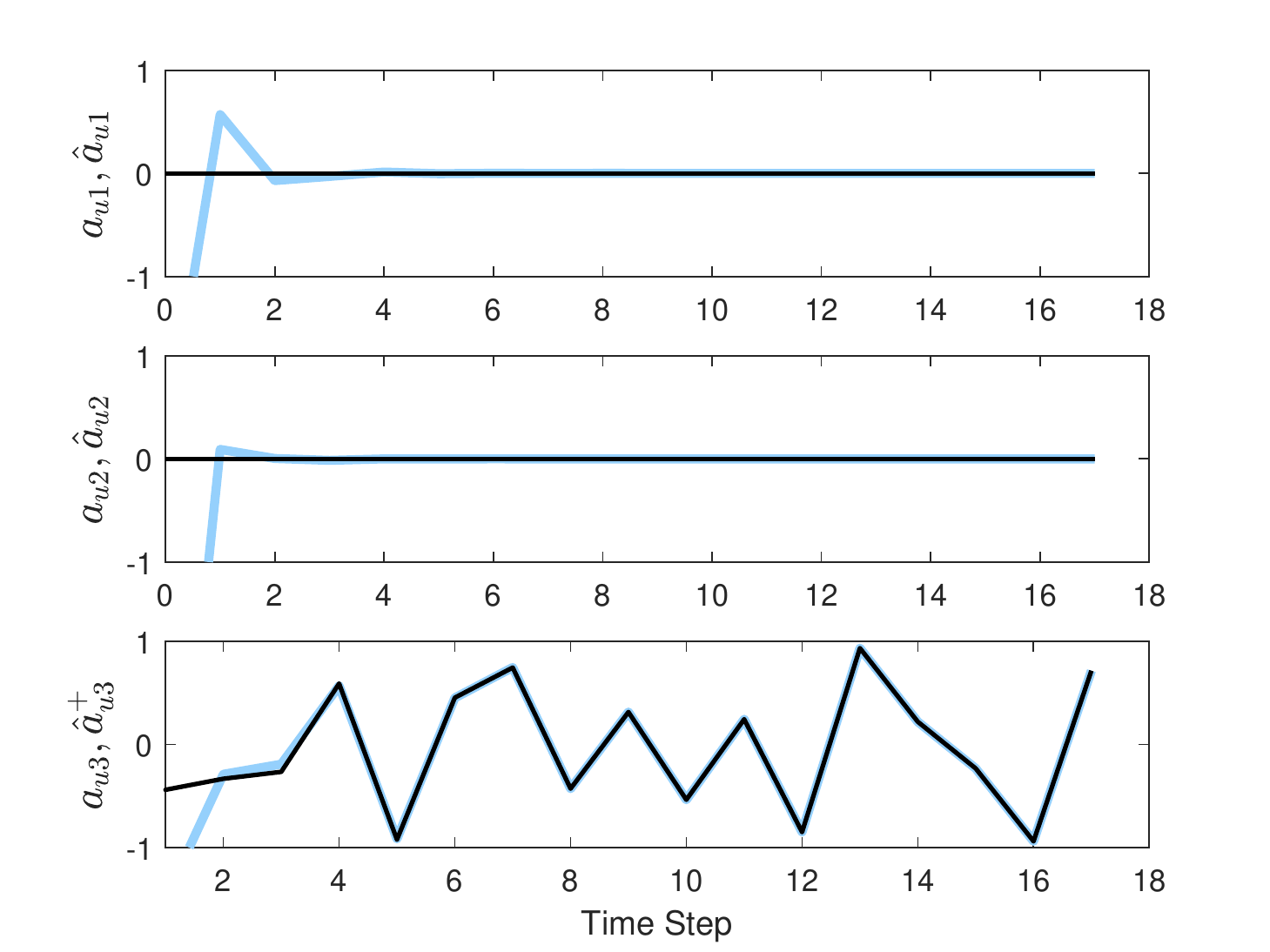}
		\caption{Estimate of $a_{u}$ when $a_{u3},a_{y2}\sim\mathcal{U}(-1,1)$. }
		\label{c332}
		\centering
	\end{figure}

\begin{figure}[t]\centering
		\includegraphics[width=0.45\textwidth]{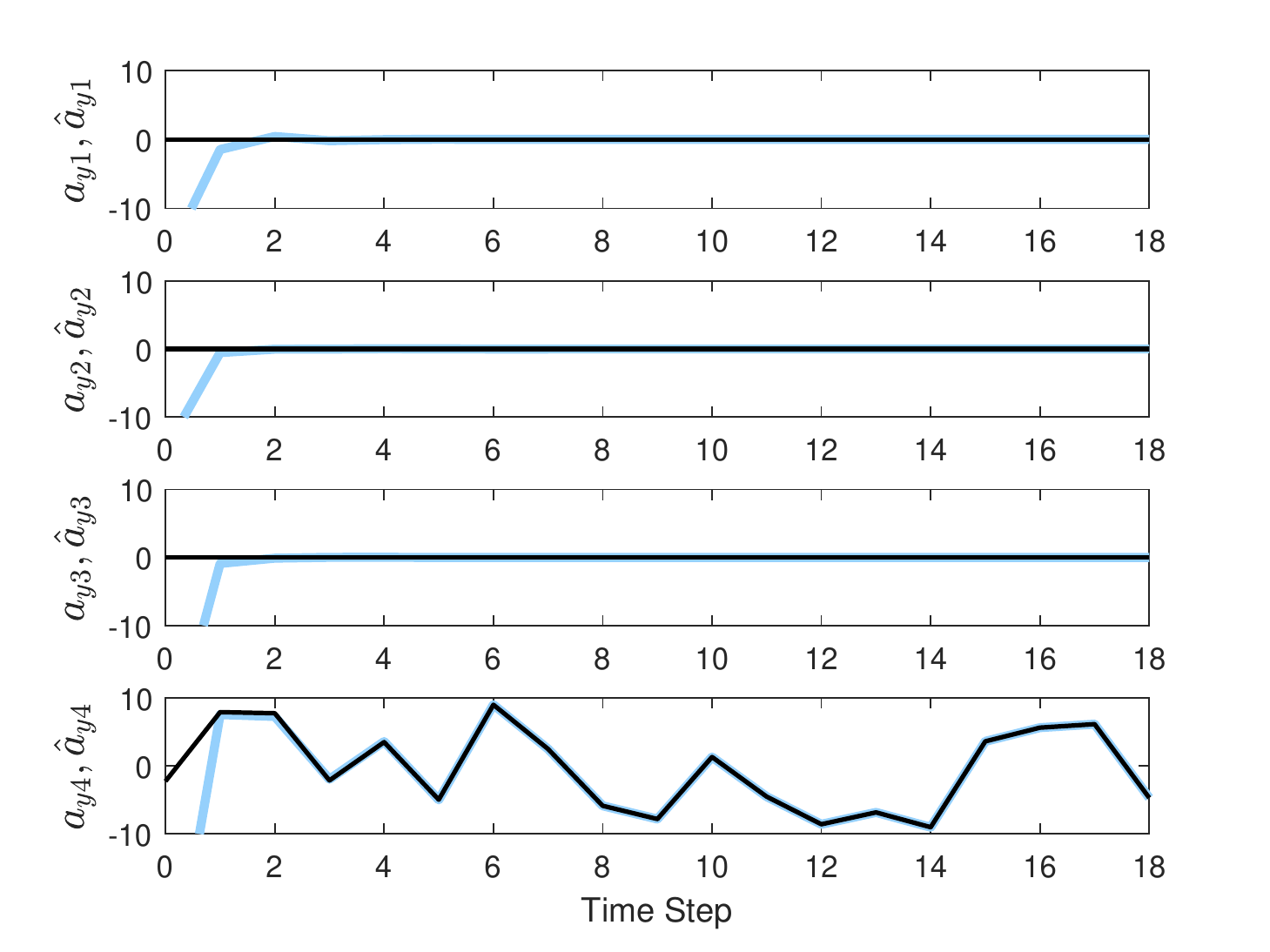}
		\caption{Estimate of $a_{y}$ when $a_{u3},a_{y2}\sim\mathcal{U}(-10,10)$. }
		\label{c333}
		\centering
	\end{figure}

\begin{figure}[t]\centering
		\includegraphics[width=0.45\textwidth]{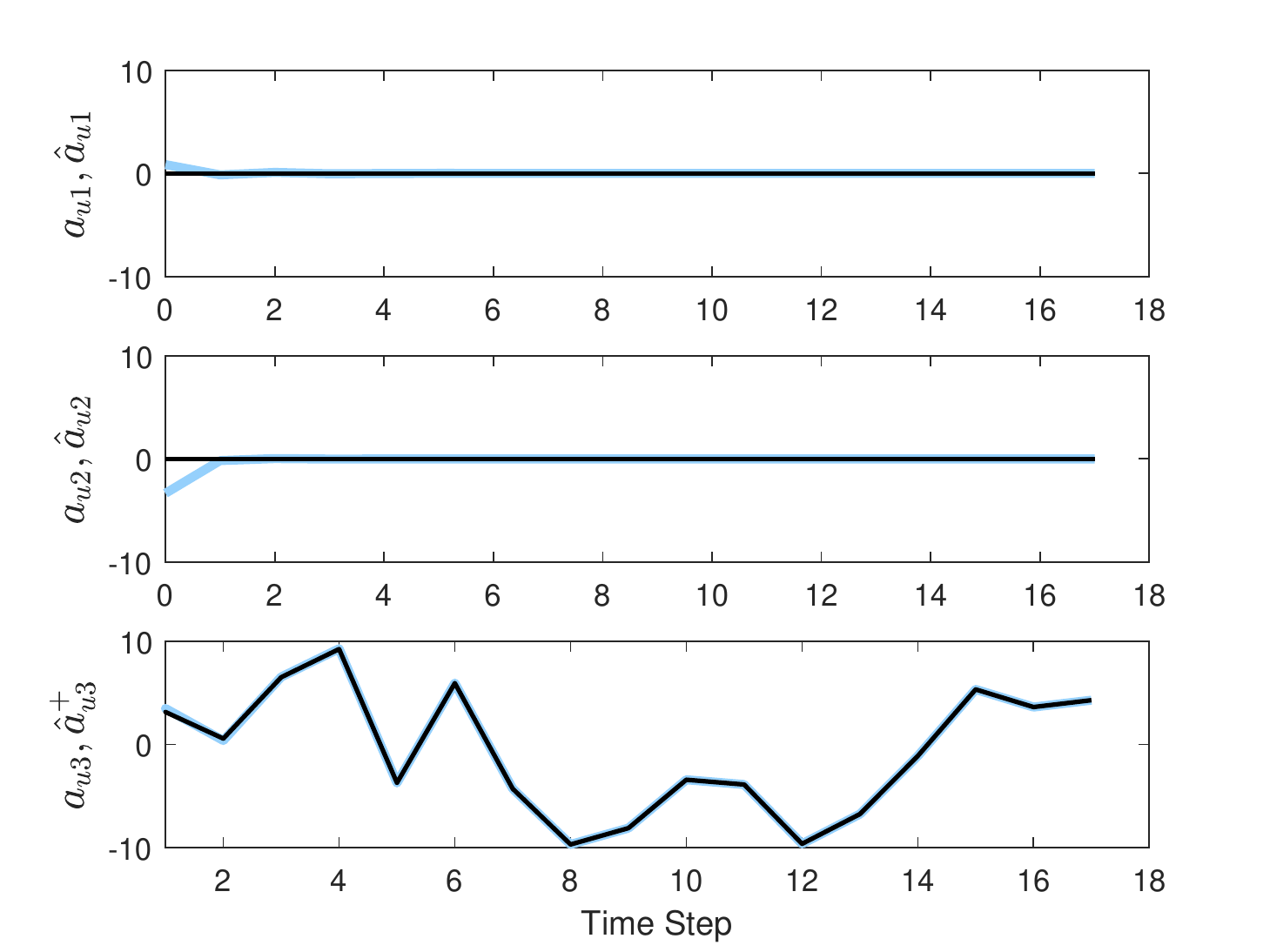}
		\caption{Estimate of $a_{u}$ when $a_{u3},a_{y2}\sim\mathcal{U}(-10,10)$. }
		\label{c334}
		\centering
	\end{figure}

	\textbf{Example 4}
	We consider model (\ref{c3e2}) in Example 2. We let $q_{1}=q_{2}=1$, $W_{u}=\left\lbrace 3\right\rbrace $, $W_{y}=\left\lbrace 2\right\rbrace $, $(u_{1},u_{2},u_{3})\sim\mathcal{U}(-5,5)$, $(a_{u3},a_{y2})\sim\mathcal{U}(-10,10)$, and $(x_{1}(0),x_{2}(0),x_{3}(0))\sim\mathcal{N}(0,1)$. We run $\binom{3}{2}\times\binom{4}{2}+\binom{3}{1}\times\binom{4}{3}=30$ partial UIOs initialized at $\hat{x}(0)=\left[ 0,0\right] ^{\top}$. We reconstruct $a_{y}$ and $a_{u}$ from (\ref{c3ry})-(\ref{c3ru}) in $19$ time-steps. The performance is shown in Figures \ref{c331}-\ref{c334}. By checking the sparsity of $a_{y}$ and $a_{u}$, actuator $3$ and sensor $2$ can be isolated as the attacked ones.
	
	\section{Conclusion}\label{conclusion}
	
	Exploiting redundancy in actuators and sensors, we have addressed and solved the problem of secure estimation and attack isolation for discrete-time nonlinear systems in the presence of (potentially unbounded) actuator and sensor attacks. We use Unknown Input Observers (UIOs) as the main ingredient for constructing an estimator capable of asymptotically reconstructing the system states and the attack signals. We use these estimates to pinpoint attacked actuators and sensors. Numerical examples are presented to illustrate the performance of our methods.
	\bibliographystyle{ieeetr}
	\bibliography{Observer1}

\end{document}